\documentclass[twocolumn]{aastex63}

\usepackage{xspace}
\usepackage{amsmath}
\usepackage{mathrsfs}

\newcommand{\Kepler}{\textit{Kepler}\xspace}

\newcommand{\Gaia}{\textit{Gaia}\xspace}

\newcommand{\Rstar}{\ensuremath{R_{\star}}\xspace} 

\newcommand{\Rjup}{\ensuremath{R_\mathrm{J}}\xspace} 
\newcommand{\Mjup}{\ensuremath{M_\mathrm{J}}\xspace}

\newcommand{\Msun}{\ensuremath{M_\odot}\xspace}

\newcommand{\absg}{\ensuremath{M_\mathrm{G}}\xspace}
\newcommand{\bprp}{\ensuremath{G_\mathrm{BP} - G_\mathrm{RP}}\xspace}

\begin{document}
%% Front Matter
\title{How Complete Are Surveys for Nearby Transiting Hot Jupiters?}

\author[0000-0001-7961-3907]{Samuel W.\ Yee}
\author[0000-0002-4265-047X]{Joshua N.\ Winn}
\author[0000-0001-8732-6166]{Joel D.\ Hartman}
\affiliation{Department of Astrophysical Sciences, Princeton University, 4 Ivy Lane, Princeton, NJ 08544, USA}

\begin{abstract}
Hot Jupiters are a rare and interesting outcome of planet formation.
Although more than 500 hot Jupiters (HJs) are known, most of them were discovered
by a heterogeneous collection of surveys with selection biases that are difficult to quantify.
Currently, our best knowledge of HJ demographics around FGK stars comes from the sample of $\approx40$ objects detected by the \Kepler mission, which have a well-quantified selection function.
Using the \Kepler results, we simulate the characteristics of the population of nearby transiting HJs. A comparison between the known sample of nearby HJs and simulated magnitude-limited samples leads to four conclusions:
(1) The known sample of HJs appears to be $\approx$\,75\% complete for stars brighter than \Gaia $G\leq10.5$, falling to $\lesssim$\,50\% for $G\leq12$.
(2) There are probably a few undiscovered HJs with host stars brighter than $G\approx10$
located within 10$^\circ$ of the Galactic plane.
(3) The period and radius distributions of HJs may differ for F-type hosts (which dominate the nearby sample) and G-type hosts (which dominate the \Kepler sample).
(4) To obtain a magnitude-limited sample of HJs that is larger than the \Kepler sample by an order of magnitude, the limiting magnitude should be
approximately $G\approx12.5$. 
This magnitude limit is within the range for which NASA's Transiting Exoplanet Survey Satellite (TESS) can easily detect HJs, presenting the opportunity to greatly expand our knowledge of hot Jupiter demographics.
~\\ % these are here because otherwise there's no space between the
~\\ % abstract and the introduction
\end{abstract}

\section{Introduction}

Hot Jupiters --- planets with masses exceeding 0.1~\Mjup and orbital periods shorter than 10 days --- are rare.
The probability that a Sun-like star has a hot Jupiter has been variously estimated as 0.4\% -- 1.2\% \citep{Mayor2011,Wright2012,Fressin2013,Petigura2018,Masuda2017}.
Despite this low probability, many of the earliest discovered planets were hot Jupiters, including 51\,Pegasi\,b \citep{Mayor1995}.
Today, hot Jupiters constitute more than $10\%$ of the sample of confirmed exoplanets.
This is because hot Jupiters are relatively easy to detect, producing strong radial-velocity (RV) signals and having conveniently short orbital periods.
They also have high transit probabilities, and when they do transit, the fractional loss of light is relatively large.

The discovery of hot Jupiters was unexpected because giant planets were supposed to form in orbits wider than a few AU, according to the prevailing core-accretion theory for giant planet formation (e.g. \citealt{Lissauer1993}, although see also \citealt{Struve1952}).
There are three categories of responses to this theoretical challenge, as reviewed by \cite{Dawson2018} and \cite{Fortney2021}.
One scenario is that hot Jupiters did form in very close orbits,
i.e., a sufficiently massive core was able to form within the inner disk despite the low surface density of solids \citep{Rafikov2006,Lee2016,Batygin2016}.
In the other two scenarios, giant planets form in wide orbits and then move inward.
The inward migration might be caused by torques from the protoplanetary disk \citep{Goldreich1980,Lin1996,Baruteau2014}, or by eccentricity excitation followed by tidal circularization \citep{Rasio1996,Fabrycky2007}.

It might be possible to figure out how often each of these processes takes place by examining the properties and statistics of hot Jupiters.
For example, migration through a disk would lead to an inner boundary for the orbital distances of hot Jupiters corresponding to the inner edge of the disk, while for high-eccentricity migration the inner boundary should occur at roughly twice the Roche radius.
Indeed, a possible pile-up of hot Jupiters at an orbital period of $\approx3$~days was first noted by \cite{Cumming1999} and \cite{Udry2003}.
\cite{Nelson2017} examined the semimajor axis distribution of the hot Jupiters discovered by the HAT and WASP surveys, arguing that it showed evidence for multiple populations, with 85\% of hot Jupiters residing in a component consistent with having formed through high-eccentricity migration, while the remaining 15\% were consistent with disk migration.

Other trends in the population of known hot Jupiters include an increasing occurrence rate with stellar metallicity \citep{Santos2004,Valenti2005}, and possibly also with stellar mass \citep{Johnson2010}, although \cite{Obermeier2016} argued that the current sample size is too small to draw a robust conclusion.
When the planet radius/period distribution is examined for periods shorter than 10 days, there appears to be a deficit of planets with radii between about 2 and 9 Earth radii, with the lower limit increasing with orbital period. This is the so-called ``hot Neptune desert'' separating hot Jupiters from smaller and mainly rocky planets \citep{Szabo2011,Mazeh2016}. 

Despite these advances, our knowledge of the demographics of hot Jupiters remains fuzzy.
Paradoxically, we have better knowledge of the demographics of smaller planets with periods ranging out to several months, because of the large and well-understood sample of thousands of such planets that resulted from NASA's Kepler mission \citep{Borucki+2016}.
In contrast, the \Kepler sample includes only 40 confirmed
and 19 candidate hot Jupiters orbiting FGK stars.
Radial-velocity surveys have discovered 44 hot Jupiters, also a relatively small sample, and
it is difficult to interpret the results because different surveys used different selection procedures, such as a preference
for high-metallicity stars, and different detection biases. Two of the most statistically well-understood radial-velocity samples are those
analyzed by \cite{Cumming2008} and \cite{Rosenthal+2021}, which contained only 12 and 14 hot Jupiters, respectively.
More than three quarters of the total observed sample of $\approx$500 hot Jupiters comes from a heterogeneous collection of wide-field ground-based photometric surveys such as WASP \citep{Pollacco2006}, HAT \citep{Bakos2004,Bakos2013}, KELT \citep{Pepper2007}, and NGTS \citep{Wheatley2018}.
While these surveys have been very successful in detecting hot Jupiters and have made important contributions to the field, they suffered from complex and severe biases due to irregular data sampling, non-stationary noise properties,
limited knowledge of the properties of the stars that were searched, and inability to perform the necessary follow-up observations of many transit candidates.
Thus, despite the large sample size, it has proven difficult to use the results of the wide-field surveys for demographics.

The \Kepler sample of hot Jupiters is probably the most homogeneous sample that is currently available.
The \Kepler telescope observed about $2 \times 10^5$ stars for 4 years, with a sensitivity high enough to be nearly 100\% complete for hot Jupiters.
The results from \Kepler provided many new insights, such as a clearer view of the connection between radius inflation and stellar irradiation \citep{Demory2011}, constraints on the frequency of nearby companion planets \citep{Steffen2012}, and the finding that the period pile-up is only evident when the sample is restricted to metal-rich stars \citep{Dawson2013}.
New questions were also raised, such as the reason for the apparent factor-of-two discrepancy between the occurrence rate of hot Jupiters from different surveys \citep{Wright2012}.
However, as noted above, the \Kepler sample includes only 40 confirmed and 19 candidate hot Jupiters orbiting FGK stars, limiting the power of statistical studies.

The work reported in this paper was undertaken to gauge the completeness of
the existing samples of hot Jupiters in the solar neighborhood, and
estimate the number and stellar host properties of the hot Jupiters that
would need to be detected in order to enlarge the statistically useful
sample from 40 to 400.
We focused on transit detection, rather than radial-velocity detection, to take
advantage of the homogeneity and completeness of the \Kepler sample of transiting planets
and to set expectations for the large number of hot Jupiters that are detectable
using data from the NASA Transiting Exoplanet Survey Satellite (TESS) mission \citep{Ricker+2015}.
Prior work by \cite{Beatty2008} had a similar goal, to predict the number of transiting hot Jupiters that would be discovered in wide-field transit surveys as a function of magnitude and galactic latitude.
Our work incorporates the knowledge we have gained since then about hot Jupiters and their properties.
We used the \Kepler sample to estimate the number of transiting hot Jupiters we should expect in a magnitude-limited sample of FGK stars (Section \ref{sec:method}), and compared this to the known sample (Section \ref{sec:results}).
Because our results suggested that the current sample is reasonably complete
down to a Gaia magnitude of 10.5, we took the opportunity to
compare this subsample of hot Jupiters with the independent
{\it Kepler} sample (Section \ref{sec:discussion}) and assess the level
of agreement in their observed properties.

\section{Completeness of the Sample of Nearby Transiting Hot Jupiters \label{sec:method}}

To gauge the completeness of the current sample of transiting hot Jupiters (HJ), we wanted to estimate how many would have been detected in a magnitude-limited survey
of nearby FGK stars.
To set expectations for more complicated models, let us start with a simple model.
In an idealized magnitude-limited transit survey of a population of identical Sun-like stars isotropically and uniformly distributed in space, the expected total number of detections is
\begin{equation}
\label{eq:simple}
    N(m_{\rm lim}) = nfp_{\rm tra} \times \frac{4\pi}{3} d_{\rm ref}^3 \times 10^{0.6(m_{\rm lim} - m_{\rm ref})},
\end{equation}
where $m_{\rm lim}$ is the limiting
apparent magnitude,
$n$ is the number density of stars,
$f$ is the fraction of stars with hot Jupiters,
$p_{\rm tra}$ is the average geometric transit probability,
and $m_{\rm ref}$ is the apparent magnitude of a Sun-like star at an arbitrarily chosen reference distance $d_{\rm ref}$.

Based on the Gaia Catalog of Nearby Stars \citep{GCNS}, a reasonable estimate for the number density of stars with absolute magnitudes between 3.5 and 6.5 (roughly the spectral types F6 through K4) is 0.007 stars/pc$^3$.
For $n$, we adopt a hot Jupiter occurrence rate of 0.6\% based on the analysis
by \cite{Petigura2018} of the \Kepler sample.
A reasonable estimate of $p_{\rm tra}$ is 0.1, corresponding to an orbital distance
of about 0.05 AU around a Sun-like star.
Further choosing $d_{\rm ref} = 116$\,pc and $m_{\rm ref} = 10$ as appropriate
for the Sun, we obtain
\begin{equation}
    N(m_{\rm lim}) \approx 25 \times 10^{0.6(m_{\rm lim} - 10)}.
\end{equation}
The current sample of hot Jupiters includes 19 with host stars
brighter than $G=10$, suggesting that it may be $\approx$\,75\%
complete down to
that magnitude. 
For $m_{\rm lim} = 8$, this formula predicts $N\approx 1.6$,
and there are 2 known hot Jupiters (HD\,209458b and KELT-11b) with host stars
with $G<8$ and spectral types between F6 and K4.
This formula also predicts that to obtain a magnitude-limited
sample of 400 hot Jupiters, the required limiting magnitude is 12.0.

We wanted to go beyond this crude approximation in order to:
\begin{enumerate}
    \item Take into account the uncertainties arising from Poisson fluctuations and the uncertainty in the hot Jupiter occcurence rate.
    \item Investigate completeness as a function of galactic latitude, to see if the practical problems associated with detecting and confirming planets in crowded star fields have led to lower completeness near the galactic plane.
    \item Account for stars of different stellar types and possible variation in hot Jupiter occurrence rate with stellar mass.
    \item Use the statistics of \Kepler transit detections directly, instead of relying on an inferred occurrence rate and a typical transit probability.
\end{enumerate}
The latter goal is complicated by the fact that \Kepler was not a magnitude-limited survey. The stars for which data are available were selected based on various criteria related to planet detectability, which depended on stellar effective temperature, radius, and the surface density of nearby stars on the sky \citep{Batalha2010}.

Our chosen method builds on similar work by \cite{Masuda2017}, who calculated the expected number of detections of transiting hot Jupiters in the globular cluster 47 Tucanae.
In short, we constructed a sample of \Kepler stars for which any transiting hot Jupiters would have been detected.
We then used the \Gaia catalog to construct a magnitude-limited sample of stars spanning the same range of colors and luminosities as the \Kepler sample --- the ``local'' sample --- and matched each local star with a star of similar color and luminosity in the \Kepler sample.
Whenever a local star was matched to a \Kepler star that hosts a transiting hot Jupiter, we assigned the local star a planet with the same properties.
We then counted the total number of transiting planets in this ``matched'' catalog, and compared it to the number of hot Jupiters actually detected in the ``local'' sample.
By repeating this process many times, we derived the statistical uncertainty in this estimate arising from Poisson fluctuations and the limited number of hot Jupiters detected by \Kepler.

The underlying premise of this method is that planet occurrence in the solar neighborhood is the same as in the \Kepler sample.
As we noted in the introduction, \Kepler hot Jupiter statistics appear to differ from those found by radial-velocity surveys by 2--3$\sigma$, which may be due to differences in the underlying stellar distribution.
Our matching procedure attempts to correct for differences in stellar population in color-magnitude space, but not directly for other factors that may affect the hot Jupiter occurrence rate, such as stellar metallicity, multiplicity, and age.
We discuss the validity of our assumptions later in Section \ref{ssec:prev_occurrence}.
In the rest of this section, we outline our data selection and matching procedure in greater detail.

\subsection{Target Selection \label{ssec:selection}}

\begin{figure*}
\epsscale{1.15}
\plotone{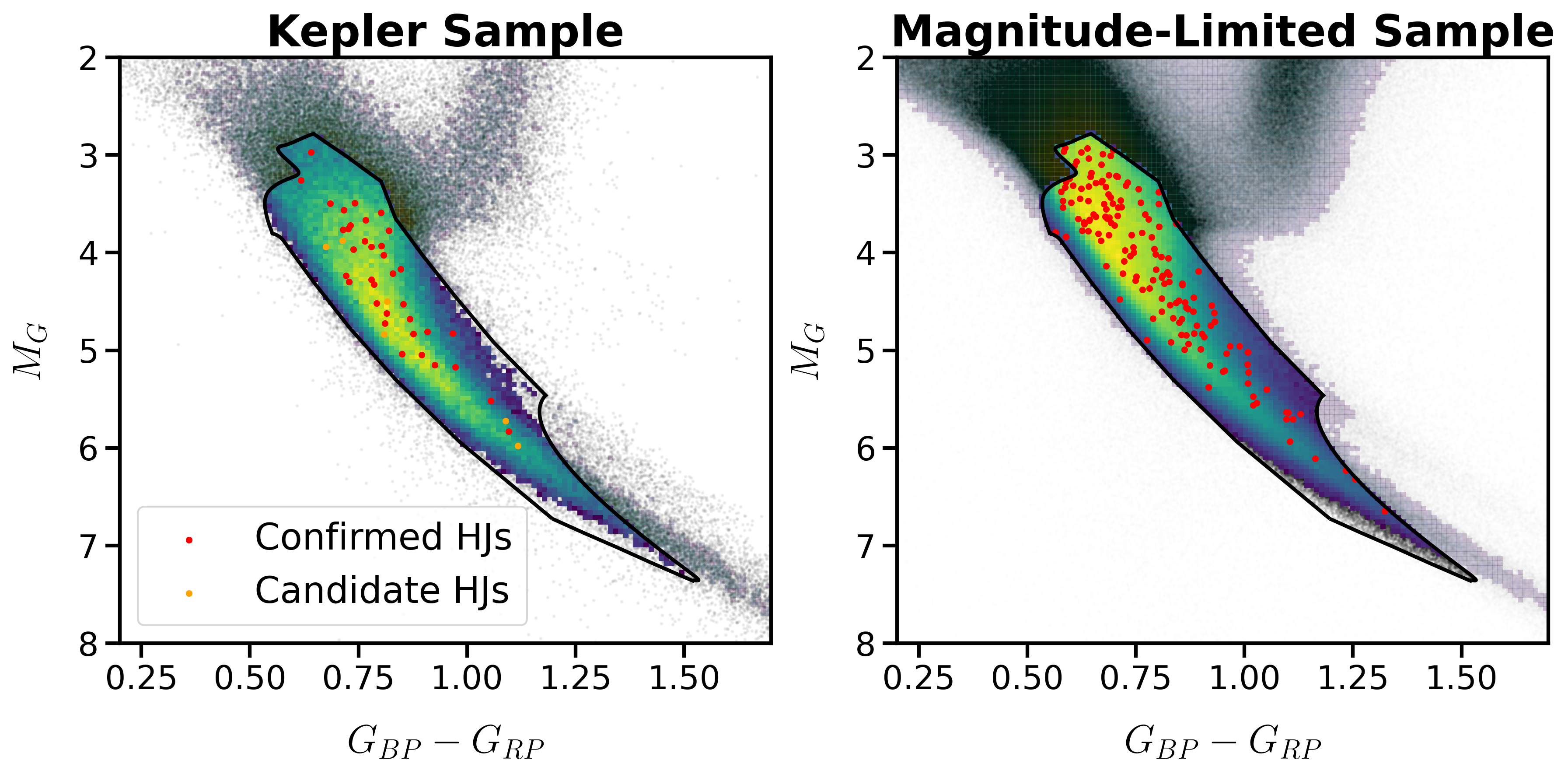}
\caption{\label{fig:cmd}
Color-magnitude diagrams for the stars in the \Kepler sample (left) and the \Gaia magnitude-limited ($G < 12.5$) sample (right).
The black boundary encloses the stars considered in our calculations, as described in Section \ref{ssec:selection}.
All magnitudes and colors have been corrected for extinction.
Red points indicate the hosts of confirmed and candidate hot Jupiters observed by \Kepler, and confirmed hot Jupiters in the magnitude-limited sample.
}
\end{figure*}

\subsubsection{Planet Properties}
The occurrence rates reported in the literature are sometimes based on different definitions of hot Jupiters. 
For our work, we consider a planet to be a hot Jupiter when it has a radius between 0.8 and 2.5 times that of Jupiter,
and an orbital period shorter than 10 days.
Using this criterion, we found $\approx$\,500 confirmed and candidate hot Jupiters when querying the NASA Exoplanet Archive\footnote{\url{https://doi.org/10.26133/NEA12}}; however, not all of these planets orbit main-sequence FGK stars.
Below, we describe our stellar selection process.

\subsubsection{Kepler Stars}

To select the stars in the \Kepler sample, we used the Gaia-Kepler Stellar Properties Catalog \citep{Berger2020}, a homogeneous set of stellar properties derived using an isochrone analysis with \Gaia parallaxes and broadband photometry.
This catalog includes almost all of the $\approx$\,200{,}000 stars observed by \Kepler, subject to cuts based on the quality of
parallax measurements and photometry from the 2MASS survey to exclude nearly equal-brightness binaries.
We obtained \Gaia EDR3 photometry \citep{GaiaEDR3,GaiaEDR3Photometry} for all the members of the catalog.
To correct the \Gaia photometry for reddening and extinction,
we used a standard extinction law $R_V = 3.1$ and the extinction coefficients for \Gaia filters derived by \cite{Casagrande2018} (Table 2), together with the $A_V$ extinctions derived by \cite{Berger2020}.
Figure \ref{fig:cmd} shows the extinction-corrected \absg, \bprp color-magnitude diagram for this sample.

We then selected FGK main sequence stars using synthetic photometry from the MESA Isochrones and Stellar Tracks \citep[MIST;][]{Choi2016,Dotter2016}.
We defined a region in the \Gaia color-magnitude diagram bounded by the zero-age main sequence (ZAMS) and terminal-age main sequence (TAMS) isochrones in MIST for stars with masses between 0.7 and 1.2~\Msun.
We used isochrones spanning initial [Fe/H] metallicities from $-0.2$ to $+0.2$, and used the union of the regions bounded by these isochrones as our final stellar selection criterion.
This cut on stellar colors and absolute magnitudes led to a sample of 112{,}203 \Kepler targets.

We also wanted to restrict our \Kepler sample to stars around which any transiting hot Jupiters would have been detected.
For each star we calculated the Multiple Event Statistic (MES),
\begin{equation} \label{eq:mes}
\mathrm{MES} = \sqrt{\frac{T_\mathrm{obs}}{P_\mathrm{orb}}} \left( \frac{R_p}{R_\star} \right)^2 \frac{1}{\sigma_\mathrm{CDPP}(T_\mathrm{tra})},
\end{equation}
where
$T_{\rm obs}$ is the total timespan of observations for a given star,
$P_{\rm orb}$ is the orbital period,
$R_p$ is the planetary radius,
$R_\star$ is the stellar radius,
and $\sigma_\mathrm{CDPP}$ is the robust root-mean-squared
Combined Differential Photometric Precision \citep{Christiansen2012}
on a timescale equal to the maximum possible duration of a transit for
a planet on a circular orbit, 
\begin{equation}
\label{eq:transit_duration}
T_\mathrm{tra,\,max} \approx 13\,\mathrm{hr}\left(\frac{P_\mathrm{orb}}{1\,\mathrm{yr}}\right)^{1/3} \left(\frac{\rho_\star}{\rho_\odot}\right)^{-1/3},
\end{equation}
where $\rho_\star$ is the mean density of the star.
The final version of the NASA \Kepler pipeline used a minimum MES threshold of $7.1$ to identify Threshold Crossing Events \citep{Twicken2016,Thompson2018}.
We computed the MES for each \Kepler target assuming $R_p=0.8\,R_{\rm Jup}$
and $P=10$\,days, the most difficult hot Jupiter to detect according
to our definition of hot Jupiters. We used the
stellar properties of \cite{Berger2020} and set
$T_{\rm obs} = (\pi/4)\,T_{\rm tra,\,max}$
to account for averaging over possible
transit impact parameters.
The value of $\sigma_\mathrm{CDPP}$ was computed for various fixed timescales based on \Kepler DR25  \citep{Twicken2016}; we used linear interpolation
to compute $\sigma_\mathrm{CDPP}$ for
our desired transit duration.

Given the calculated MES values for the detection of a hot Jupiter around each target, we excluded those for which MES is less than 7.1.
We also excluded stars for which the observation duration $T_\mathrm{obs}$ was less than
30~days, given that at least three transits needed
to be observed for a secure detection.

Only 13 of the stars in our \Kepler sample were excluded by these criteria. Increasing the MES threshold from 7.1 to $10$ or $17$ changed the number of stars in our final sample by less than one percent.
Thus, this exercise served to confirm that
\Kepler could have detected a transiting hot
Jupiter around essentially all of
the stars it observed during its 4-year primary mission.
This reinforces our notion that the hot Jupiters detected by \Kepler represent the most complete and well-understood sample of such planets currently available.

Around the stars meeting our selection criteria, \Kepler detected a total of 40 confirmed hot Jupiters, and 19 candidate transit signals
for which the reported light-curve properties are consistent
with hot Jupiters.
We excluded planets with grazing transits (impact parameters $>$\,0.9) because of their
lower detectability, larger uncertainties in
planet radius and other parameters, and
higher likelihood of being false positives.
This left us with a sample of 36 confirmed and 6 candidate hot Jupiters.
Each of the 6 candidates was assigned
a False Positive Probability (FPP) by \cite{Morton2016}.
By assigning each of the confirmed
planets a weight of 1.0, and each
of the candidates a weight of
$1 - $~FPP,
we arrived at an effective sample size of 41 transiting hot Jupiters drawn from a sample
of 112{,}203 stars.
We will refer to this \Kepler sample
by the symbol $\mathcal{S}_K$.

\subsubsection{Magnitude-Limited Sample}
We then constructed a magnitude-limited sample using data from \Gaia EDR3 \citep{GaiaEDR3}. We queried the \Gaia
archive for all stars brighter than $G < 12.5$ to obtain the photometric and astrometric observations, using a standard quality cut on the parallax ($\varpi / \sigma_\varpi > 5$)
to remove suspect data.
We also used the geometric distances from \cite{Bailer-Jones2021} to compute absolute $G$-band magnitudes.
Although most of these bright stars are relatively nearby and do not suffer significant dust extinction, we nonetheless corrected their \Gaia $G$, $G_\mathrm{BP}$, and $G_\mathrm{RP}$ magnitudes using the \texttt{mwdust} Python package \citep{Bovy2016}.
In particular, we used the \texttt{Combined19} dust map, which combines the maps from \cite{Green2019}, \cite{Marshall2006}, and \cite{Drimmel2003} to provide full sky coverage.
The majority of our stellar sample received only small corrections, with $90\%$ of stars having $E(G_\mathrm{BP} - G_\mathrm{RP}) < 0.15$.

To select main sequence FGK stars, we applied the same cut in the extinction-corrected
color-magnitude diagram (\absg versus \bprp)
as we did for the \Kepler sample.
We did not make any further cuts on astrometric fit quality, out of concern
that hot Jupiters may be preferentially associated with stars with wide-orbiting companions \citep[e.g.,][]{Ngo2016} which would affect the quality of the \Gaia astrometric fits (\citealt{Belokurov2020}, although see also \citealt{Moe2020}).
This procedure yielded 1{,}073{,}225 stars in our magnitude-limited ``local'' sample, which we denote by the symbol $\mathcal{S}$.
According to the NASA Exoplanet Archive,
there are 154 transiting hot Jupiters known to exist around the stars in this sample.

The right panel of Figure \ref{fig:cmd} shows the color-magnitude diagram for the stars and hot Jupiters in this sample.
The two stellar populations are different:
\Kepler target stars were chosen to maximize the number of small planets that could be detected \citep{Batalha2010}, leading to a dominance by G-dwarfs;
meanwhile, the Gaia magnitude-limited sample is dominated by early F stars because they are
more luminous and can be seen to a greater distance at fixed apparent magnitude (Malmquist bias).
If the hot Jupiter occurrence rate varies according to stellar type, then our matching procedure should account for these differences. This was the motivation for the process described in the following section.

\subsection{Matching Procedure \label{ssec:matching}}

\begin{figure}
\epsscale{1.15}
\plotone{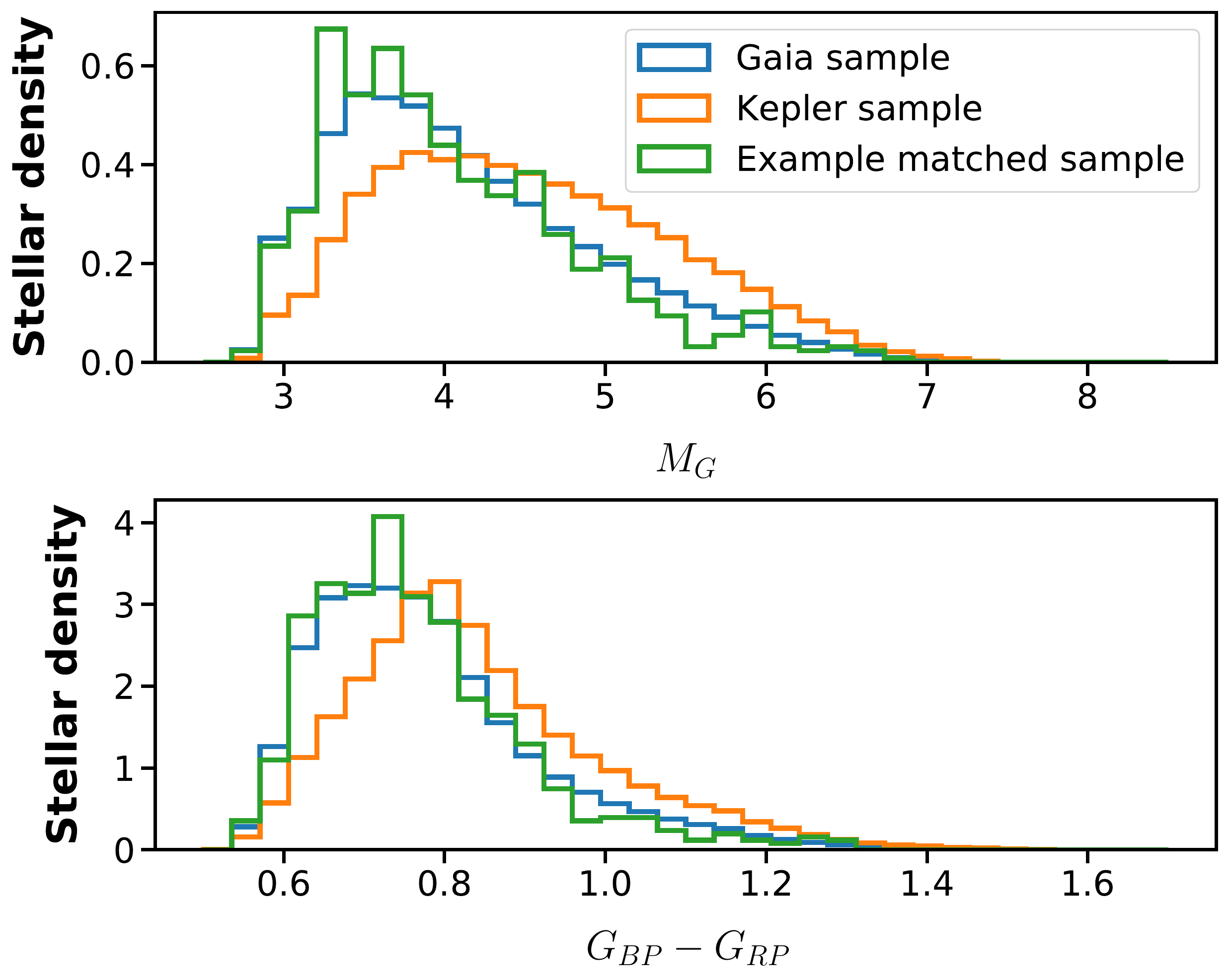}
\caption{\label{fig:matched_histogram}
Absolute magnitude and color distributions of the local sample $\mathcal{S}$,
the \Kepler sample $\mathcal{S}_K$, and an example of a matched sample.
Our nearest-neighbors matching procedure ensures that the matched sample has stellar properties similar to the stars in the local sample.
}
\end{figure}

We performed our matching procedure to generate a synthetic catalog of transiting hot Jupiters around the stars in $\mathcal{S}$.
First, we drew stars, with replacement, from $\mathcal{S}_K$, to generate a new sample $\tilde{\mathcal{S}}_K$ that has the same number of members as $\mathcal{S}$.
This step accounted for the Poisson fluctuations in the number of planets in the \Kepler sample.
We then associated each star in $\mathcal{S}$ with a star in the resampled set, $\tilde{\mathcal{S}}_K$.
To account for the differences in the underlying stellar populations of the two samples, and any possible variation in hot Jupiter occurrence with stellar type, we wanted to match the stars in $\mathcal{S}$ with stars of similar spectral types in $\tilde{\mathcal{S}}_K$.

We defined a metric in color-magnitude space,
\begin{equation}
    d = \sqrt{
    \left( \frac{\Delta{\rm mag}}{\sigma_{\rm mag}} \right)^2 + 
    \left( \frac{\Delta{\rm color}}{\sigma_{\rm color}} \right)^2
    },
\end{equation}
where $\Delta$mag is the difference in absolute
magnitude,
$\Delta$color is the difference in the \bprp color,
and $\sigma_{\rm mag}$ and $\sigma_{\rm color}$
are the standard deviations of the distributions of absolute
magnitude and color of the entire sample.
Then, for each star in $\mathcal{S}$, we drew a proposed match in $\tilde{\mathcal{S}}_K$, and accepted this proposal
with probability $P(d) \propto e^{-3d}$.
Whenever a proposed match was rejected, we proposed a new matching star from $\tilde{\mathcal{S}}_K$ until all stars in $\mathcal{S}$ were successfully matched.

The choice of 3 as the factor in the exponent of $P(d)$ was made
after some experimentation to balance two effects: if set too small, the resulting distribution would not resemble the target distribution; too high, and the sparsity of hot Jupiter hosts in the \Kepler sample would mean some stars in $\mathcal{S}$ would never be matched with a planet-hosting star in $\mathcal{S}_K$.
We arrived at 3 by computing the distances between each \Kepler hot Jupiter hosts and its five closest counterparts, and fitting the resulting distribution of distances with an exponential distribution.
In this way, all stars in $\mathcal{S}$ had a nonzero probability of being matched with a hot Jupiter host in $\mathcal{S}_K$.

This process generated a ``matched'' catalog of stars, denoted $\mathcal{S}_M$, which is comprised of real \Kepler stars with a distribution in color-magnitude space that is similar to that of the local magnitude-limited sample.
Figure $\ref{fig:matched_histogram}$ compares a single realization
of a matched catalog to the distribution of stars in the Gaia sample $\mathcal{S}$, illustrating the desired agreement in stellar properties.

We then assigned each transiting hot Jupiter around a star in $\mathcal{S}_M$ to its matched counterpart in $\mathcal{S}$.
Given that each star is only matched to a similar star, any corrections for changes in transit probability due to differing stellar densities are minimized, and were neglected in our subsequent calculations.
The end result was a synthetic catalog of transiting hot Jupiters around a magnitude-limited sample of nearby stars from \Gaia, based on their occurrence statistics in the \Kepler sample.
We repeated this process to generate 1000 realizations of this catalog in order to estimate sampling uncertainties in the results.

\section{Results} \label{sec:results}

\begin{figure}
\epsscale{1.15}
\plotone{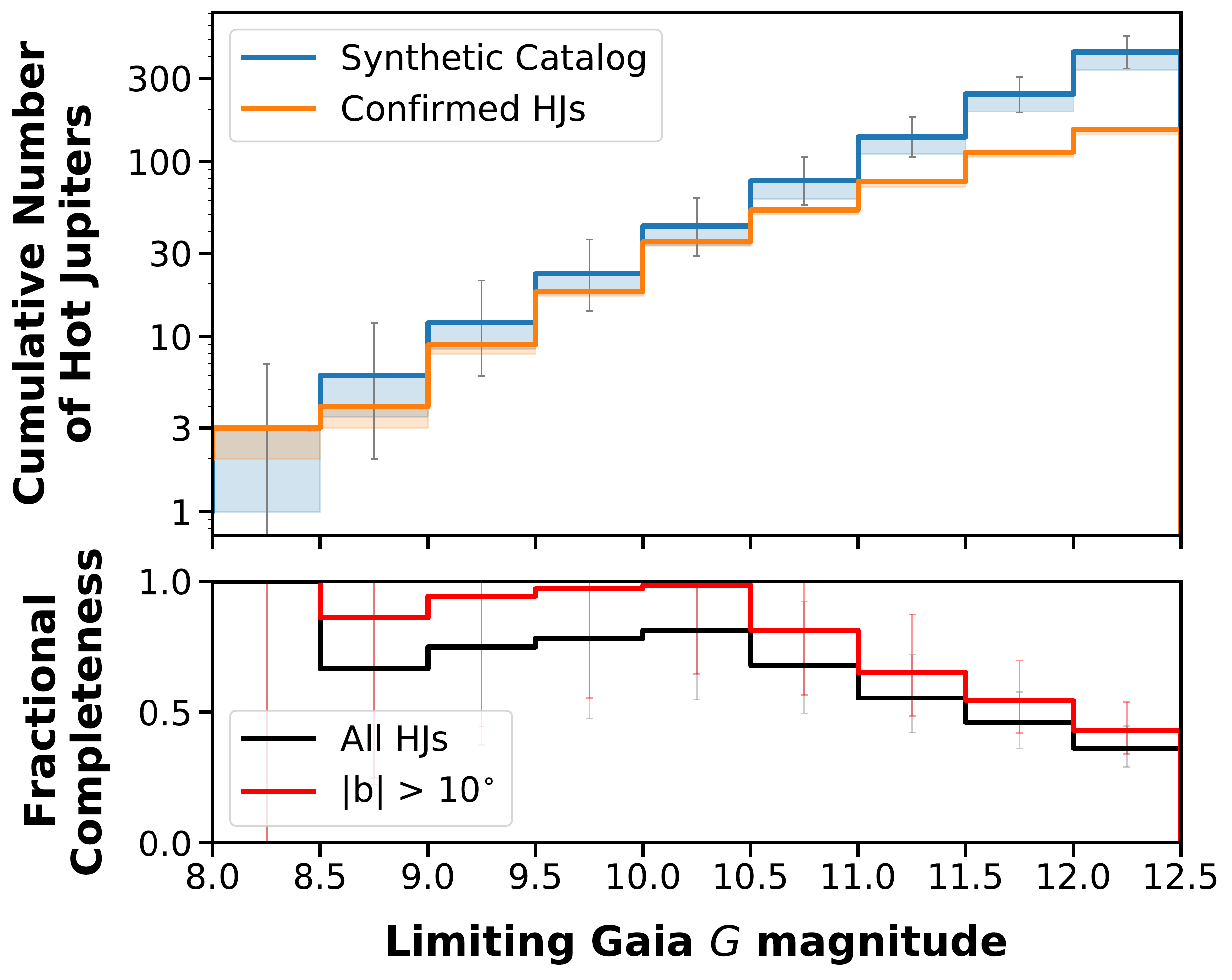}
\caption{\label{fig:hj_mag_dist}
{\bf Top}: Cumulative number of non-grazing transiting hot Jupiters as a function of limiting apparent magnitude,
based on simulated \Kepler-matched samples of nearby stars (blue)
and the actual collection of confirmed hot Jupiters around
the same stars (orange).
Error bars represent sampling uncertainties.
Shaded blue regions represent the expected contribution
of undiscovered hot Jupiters located within 10$^\circ$
of the galactic plane.
Based on the simulations, a complete survey of $G < 12.5$
FGK stars should contain $\approx$\,400 transiting hot Jupiters.
\\
{\bf Bottom}: Estimated completeness of surveys for
nearby hot Jupiters. The black histogram is for
all stars, and the red histogram is for
stars more than 10$^\circ$ from the galactic plane.
Based on this comparison,
the known sample of hot Jupiters is about 75\%
complete down to $G < 10.5$, falling to $50\%$ down
to $G < 12.5$.
}
\end{figure}

The results from this matching process are shown in Figure \ref{fig:hj_mag_dist}.
Based on these results,
we expect that a complete survey of nearby FGK stars brighter than $G < 12.5$ will contain $424^{+98}_{-83}$ transiting hot Jupiters with impact parameters $b < 0.9$ --- assuming that the \Kepler hot Jupiters are representative of the nearby population of hot Jupiters.
For comparison, 154 hot Jupiters have actually been confirmed around such stars, leaving several hundred more hot Jupiters undiscovered around nearby stars.
Figure \ref{fig:hj_mag_dist} also shows that the
apparent magnitude distribution
of known hot Jupiter hosts is consistent with being 75\% complete down to a limiting \Gaia $G$-band magnitude of 10.5.
For $G<12.5$, the estimated completeness
falls to $36^{+8}_{-7}\%$.

Recall that the simpler calculation presented at the beginning of Section \ref{sec:method} predicted that 400 hot Jupiters could be found down to a limiting magnitude of $G < 12$.
Our sample-matching procedure suggests that the actual limiting magnitude needs to be 12.4 to reasonably expect to detect 400 transiting hot Jupiters.
The differences in the results are due to the simplifying assumptions that were made in the earlier calculation as well as our restriction in the sample-matching procedure
that the planets need to show non-grazing transits.
We can correct for the latter effect by multiplying the expected number of hot Jupiters in our simulated samples by a factor
of $1/0.9$, as would be appropriate for isotropically oriented
planetary orbits.  This simple correction leads to an
expectation of $472^{+110}_{-92}$ transiting hot Jupiters
for an apparent magnitude limit of $G=12.5$.

\begin{figure}
\epsscale{1.15}
\plotone{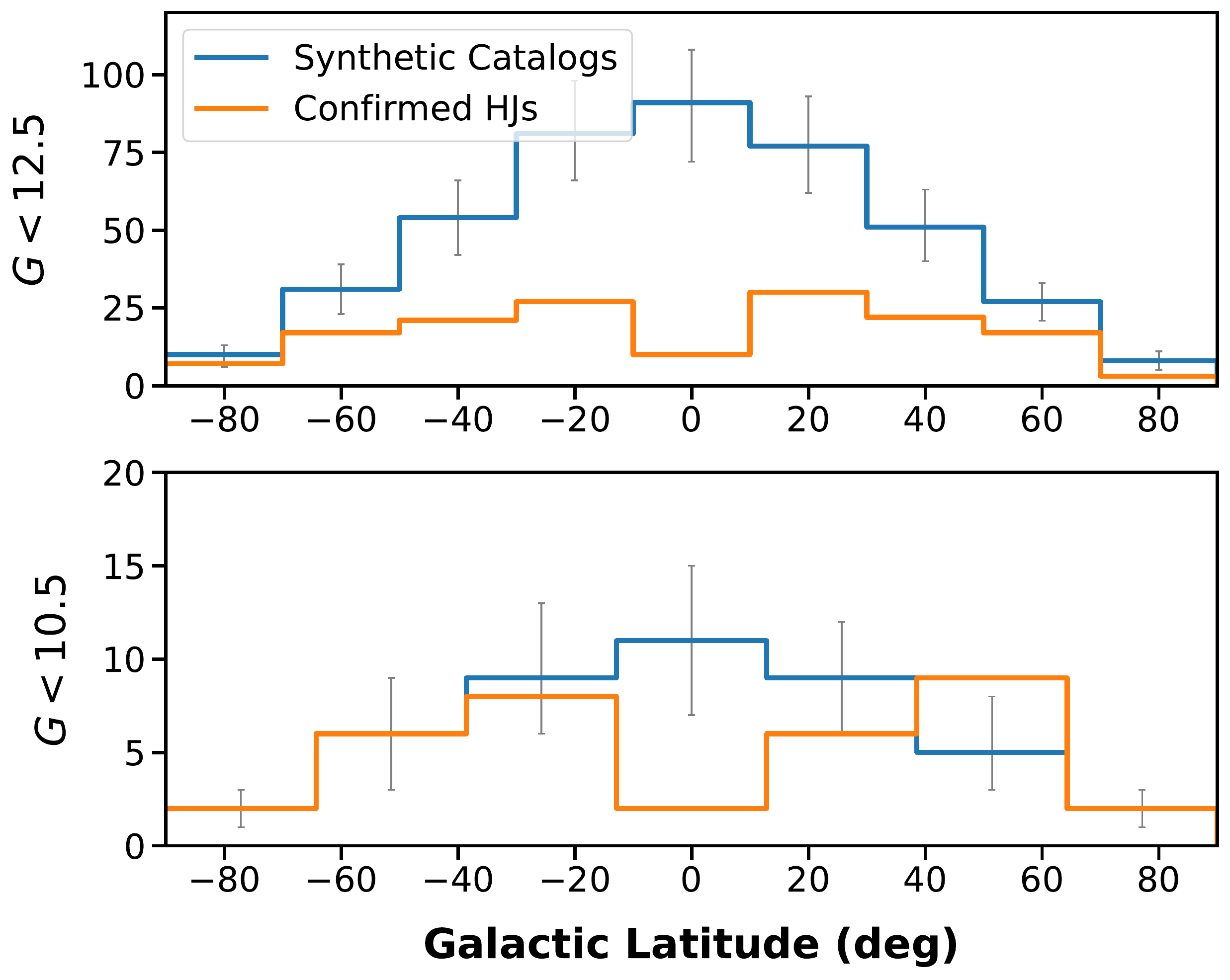}
\caption{Galactic latitude distribution of the expected number of hot Jupiters (blue), and the currently known hot Jupiters (orange), for two different apparent magnitude limits:
$G=12.5$ (top) and $G=10.5$ (bottom).
Ground-based transit surveys have tended to avoid the galactic plane due to the difficulty of achieving precise photometry
in crowded star fields.
\label{fig:glat_distribution}}
\end{figure}

\begin{figure}
\epsscale{1.15}
\plotone{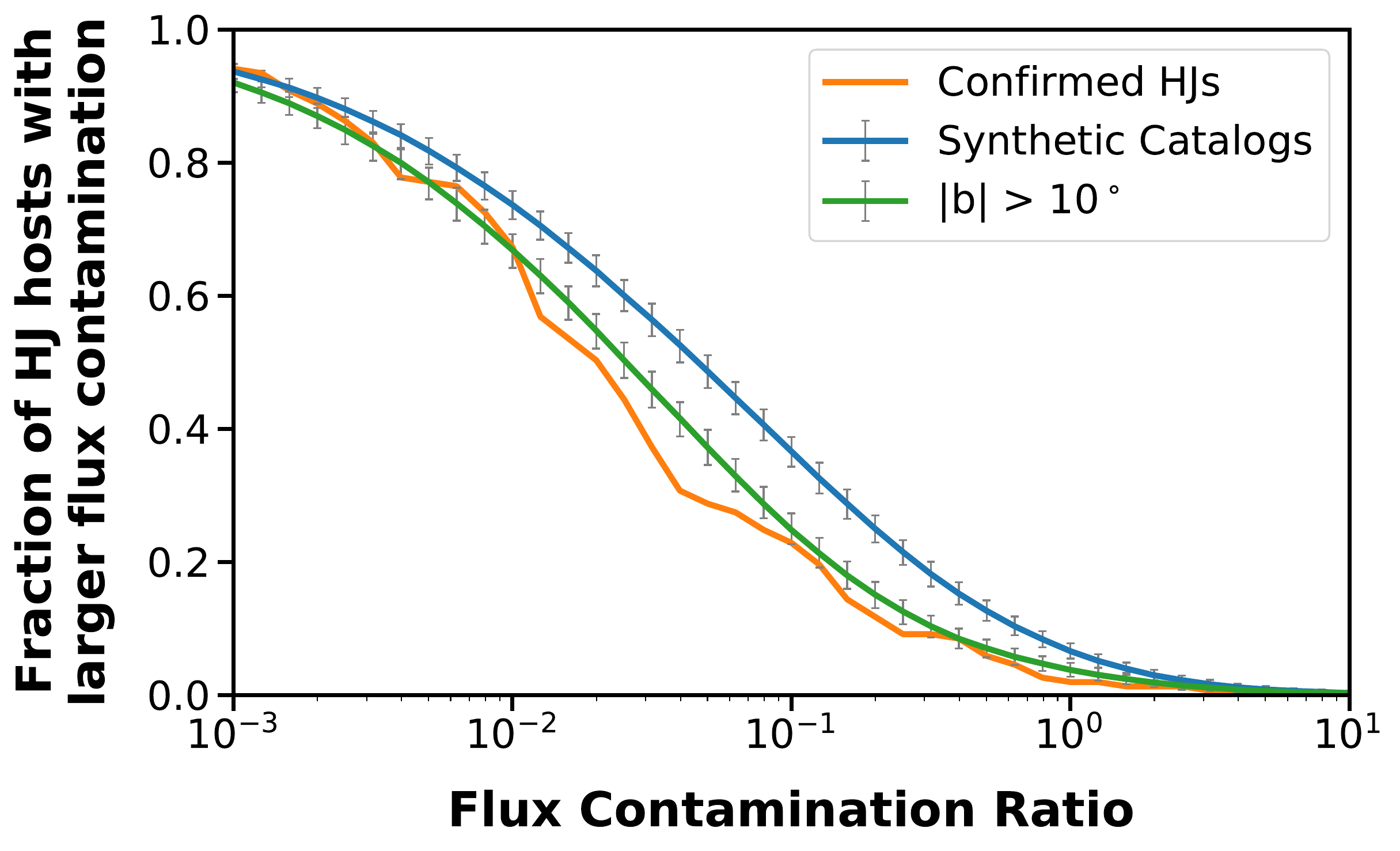}
\caption{The fraction of hot Jupiter hosts with flux contamination ratios (a measure of the crowdedness of the star field) larger than the value of the x-axis, for the stars in our synthetic catalogs (blue) and confirmed hot Jupiter hosts (orange), in both cases with an apparent magnitude limit of $G=12.5$.
The confirmed hot Jupiter hosts are biased toward lower flux contamination ratios and lower crowdedness.
If we exclude stars close to the galactic plane (green line), the distribution for the synthetic catalogs is more similar to the confirmed population.
\label{fig:cont_ratio}}
\end{figure}

One of the simplifying assumptions that was made in the basic calculation leading to Equation~\ref{eq:simple} was that stars are distributed isotropically in space around the Sun.
In reality, the number density of stars varies strongly as a function of galactic latitude, and the number density of stars also affects the detectability of transiting planets.
The crowded star fields near the galactic plane lead to various practical difficulties.
Having multiple stars within a
photometric aperture reduces the amplitude of the transit
signal and increases the noise level due to Poisson fluctuations
and variations in the point-spread function.
A separate issue is that the incidence of ``false positive'' signals
is higher in crowded fields, due to the higher density of background
eclipsing binaries whose eclipses can be mistaken for planetary transits.
Furthermore, the radial-velocity observations and other spectroscopic
follow-up observations that are necessary for planet confirmation
are made more difficult in the presence of nearby bright stars.

Figure \ref{fig:glat_distribution} shows that for a limiting magnitude of 10.5, most of the ``missing'' hot Jupiters
are probably to be found within $10^\circ$ of the
galactic plane.
This strip at low galactic latitudes subtends 17\% of the
sky and is home to $21\pm 2\%$ of hot Jupiters in our simulated samples, but it only contains 6.5\% (10 out of 154) of the sample of known hot Jupiters.
If we restrict ourselves only to stars with $|b| > 10^\circ$, then the estimated
completeness of the $G < 10.5$ magnitude-limited sample of hot Jupiters rises to $99^{+41}_{-33}\%$.
Our simulations also suggest that there are a handful (1--4) of
hot Jupiters orbiting stars brighter than 10th magnitude waiting
to be discovered by those without fear
of treading within the galactic plane.

To explore the practical effects of crowdedness
on finding and confirming planets, we examined
the TESS Candidate Target List \citep[CTL;][]{Stassun2018,Stassun2019}.
This catalog includes for each star an estimate
of the ``flux contamination ratio,'' defined as
the estimated fraction of the total flux within
3.5$'$ (10 TESS pixels) that
comes from neighboring stars.
While the search radius of 3.5$'$ is larger than the typical size of photometric apertures
used in ground-based surveys, we expect the TESS flux
contamination ratios are representative of the
degree of crowding.
We also note that the CTL flux contamination ratio was only computed based on
nearby stars resolved by Gaia DR2, and so will not include the dilution
from companions closer than $\lesssim 1"$.

Figure \ref{fig:cont_ratio} shows the distribution of 
flux contamination ratios of the known hot
Jupiter hosts and for our synthetic catalogs with an apparent magnitude limit of 12.5.
As expected, the comparison shows that the known hot Jupiter population is biased toward lower flux contamination ratios (reduced crowding).  
While $< 5\%$ of confirmed hot Jupiter hosts have contamination ratios of 0.5 and greater, almost three times that fraction of hosts in our synthetic catalogs have such large ratios.
When we exclude the synthetic catalog stars close to the galactic plane ($|b| < 10^{\circ}$), their distribution more closely matches that of the confirmed planets.
Nonetheless, most stars have relatively low flux contamination of $< 0.2$, which seems too small for the signal dilution and increased Poisson noise to cause much damage to hot Jupiter detection.
Instead, we suspect that the main factors are the higher incidence of false
positives and the difficulty of follow-up observations.
Assuming that these practical problems are not easily solved,
and therefore excluding all stars within 10$^\circ$ of the galactic
plane from all the samples, we estimate
that a complete sample of FGK stars
with $G < 12.5$ would contain $334^{+82}_{-70}$ transiting hot Jupiters,
as compared to the 144 known hot Jupiters in the currently known sample.

The earlier work of \cite{Beatty2008} presented an analytic framework for predicting the yield of transit surveys, including the effects of galactic structure and the window function for ground-based surveys.
They estimated $\approx 80$ HJs with orbital period $P < 5$~days would be found in a magnitude-limited survey of Sun-like stars down to $V \leq 12$.
Indeed, when we restrict our synthetic catalogs to $P < 5$~days and $V \leq 12$, we obtain an expected yield of $89^{+19}_{-18}$ planets.
This good agreement is in part due to \cite{Beatty2008}'s use of occurrence rates from the OGLE-III survey \citep{Gould2006}, which found a hot Jupiter occurrence rate of $0.45\%$, similar to the low occurrence rate found by \Kepler (see Section \ref{ssec:prev_occurrence}).
Our work differs from their analytic study by performing a numerical matching of stars in the all-sky and \Kepler catalogs, accounting for possible variation of occurrence rates with stellar and planet properties, and producing synthetic catalogs of transiting planets.

\section{Discussion \label{sec:discussion}} 

\subsection{Comparison of Possibly Complete Samples \label{ssec:complete_samples}}

\begin{figure}
\epsscale{1.15}
\plotone{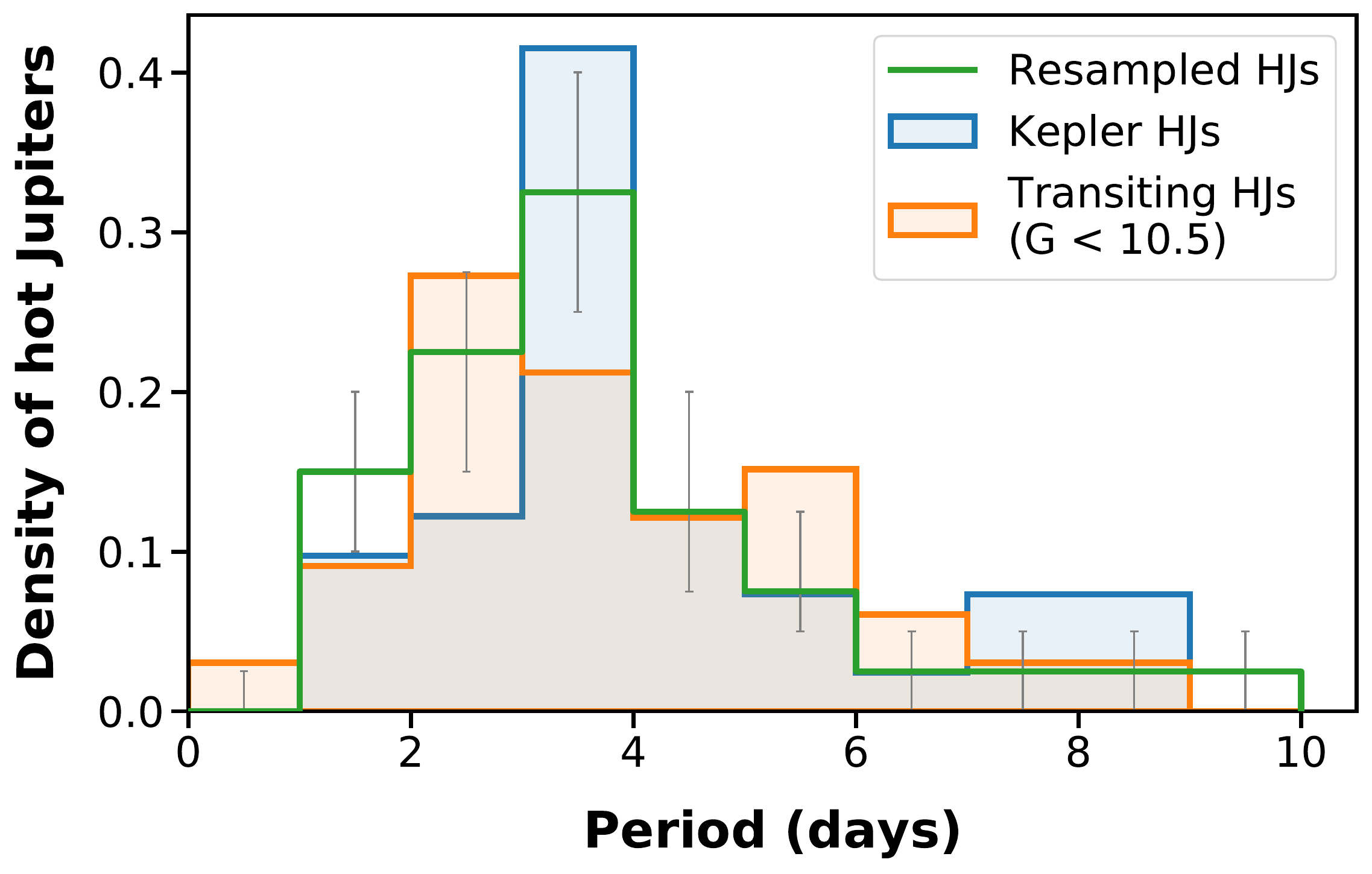}
\plotone{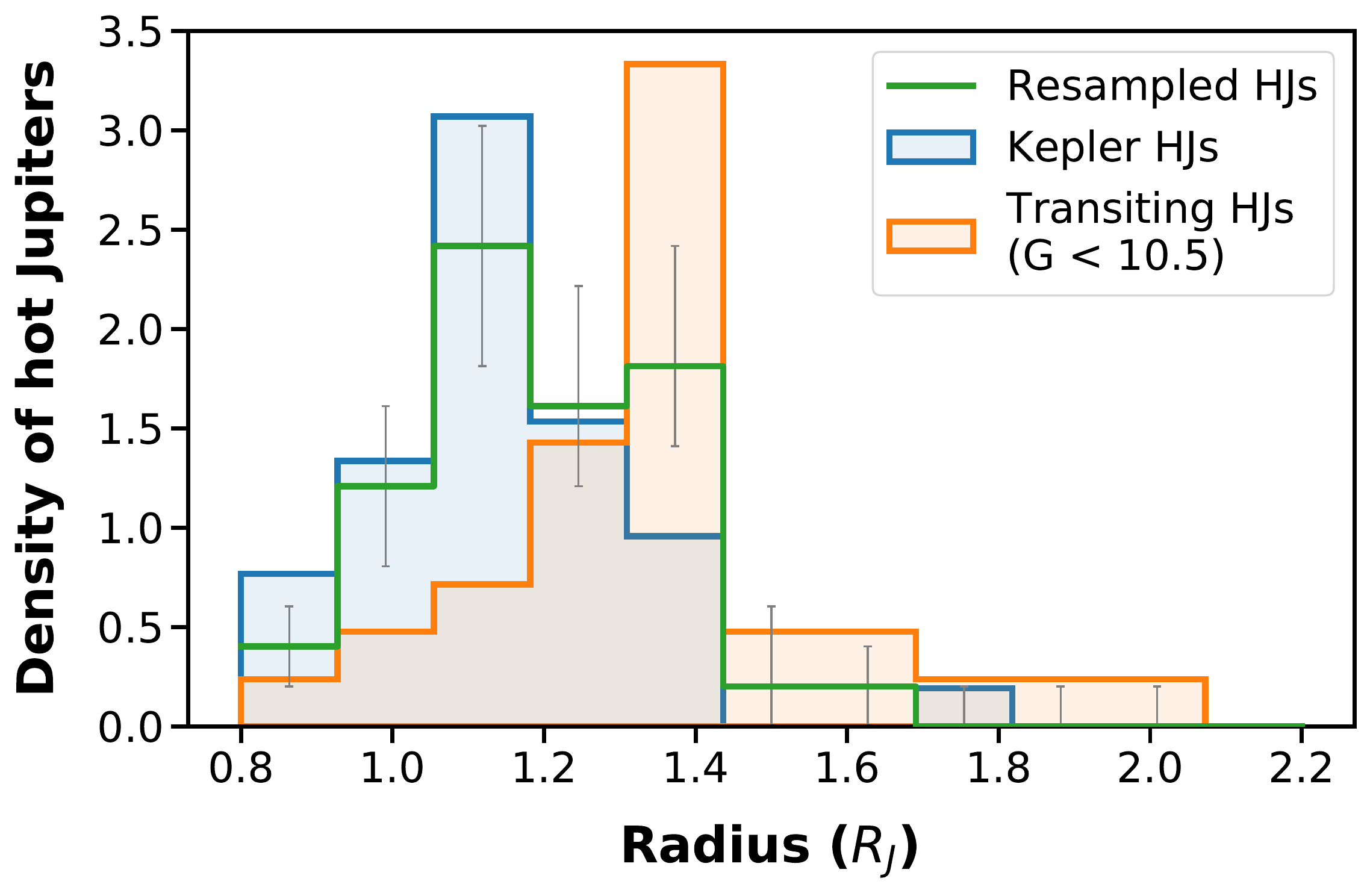}
\caption{Period and radius distributions of transiting hot Jupiters in the \Kepler (blue) and magnitude-limited sample (orange).
The two populations appear to have differing properties, but this discrepancy is reduced when we resample the transiting HJ population to match the stellar distribution of \Kepler targets (green).
The error bars on the green histogram reflect the 1-$\sigma$ widths of the distribution based on 1000 resampled catalogs.
\label{fig:period_rad_comparison}}
\end{figure}

\begin{figure}
\epsscale{1.15}
\plotone{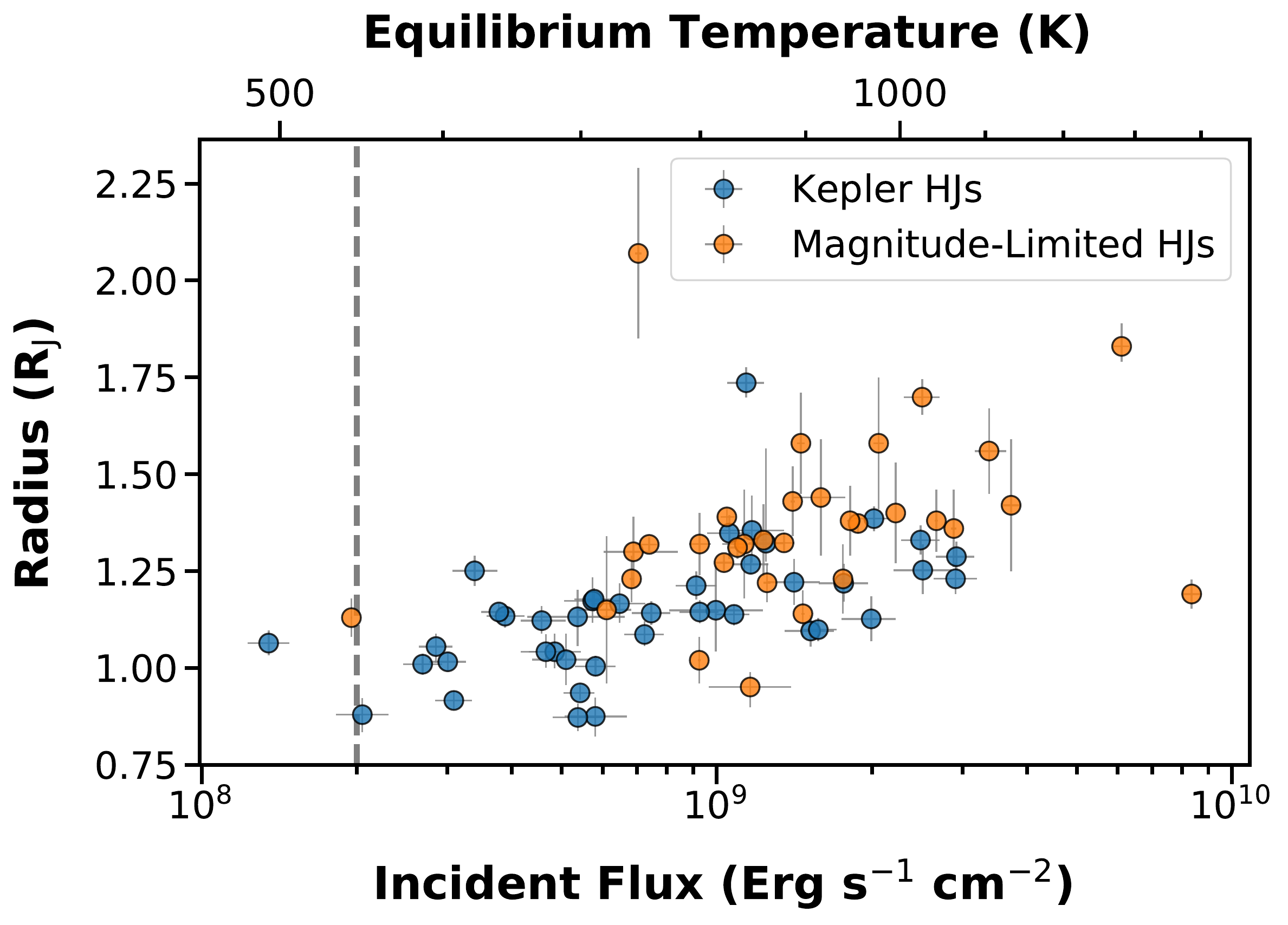}
\caption{Radius versus
stellar irradiation flux,
for the hot Jupiters in the \Kepler
and magnitude-limited samples.
The planets in the magnitude-limited sample tend to orbit hotter stars and have larger sizes than the planets in the \Kepler sample.
\label{fig:hj_insolation}}
\end{figure}

Based on the comparison between the apparent magnitude distribution of the hosts
of known hot Jupiters and the distribution one would expect based on
the incidence of transiting hot Jupiters in the \Kepler
survey, we are currently aware of about 40\% of the
hot Jupiters with host stars
brighter than $G=12.5$.
However, for a magnitude limit of $G = 10.5$, the total number of known
hot Jupiters is consistent with the expected number, once we exclude
the region of the sky within 10$^\circ$ of the galactic plane.
Our \Kepler-matching procedure predicts $33\pm7$ transiting hot Jupiters, in
agreement with the 33 transiting hot Jupiters that have been found.
This suggests that we might already be in possession of a nearly complete
magnitude-limited sample of hot Jupiters down to $G=10.5$.

Thus, we took the opportunity to compare this ``possibly complete''
subset of hot Jupiters --- which should be less biased in planet properties
than the known hot Jupiter population as a whole --- with the sample of
42 \Kepler hot Jupiters.
Figure \ref{fig:period_rad_comparison} compares the
period and radius distributions of both of these real
samples of hot Jupiters.
There are some hints of possible differences in these distributions, despite
the fact that we expect both samples to be nearly complete.
The period distribution of \Kepler hot Jupiters exhibits a pile-up
at $\approx\,3$~days, as first noted 
by \cite{Latham2011} in an early catalog of
Kepler Objects of Interest.
Any such peak does not seem as pronounced in the magnitude-limited sample of hot Jupiters.
However, visual inspection can be misleading.
A two-sided Kolmogorov-Smirnov (K-S; \citealt{Kolmogorov1933,Smirnov1948}) test cannot reject the hypothesis that the two distributions
are drawn from the same distribution ($p=0.3$). The relatively small number of planets in both samples (33 and 42 for the magnitude-limited and \Kepler samples respectively) limits the statistical power
of the comparison.

We also compared the radius distributions of the two populations, and this time, the K-S test
does reject the null hypothesis ($p= 2\times10^{-5}$).
The magnitude-limited sample contains a higher proportion of planets larger
than 1.2\,$\Rjup$ than the \Kepler sample.
At least part of this discrepancy may arise from the different distributions
of stellar types within the two samples. The \Kepler sample has a higher
fraction of G-type stars and a lower fraction of F-type stars
than the the magnitude-limited sample (Figure \ref{fig:cmd}).
Thus, the hot Jupiters in the magnitude-limited sample are subject to higher
levels of stellar irradiation (Figure \ref{fig:hj_insolation}),
which has been shown to be associated with larger planetary radii
\citep[see, e.g.,][]{Fortney2007,Demory2011}.

To investigate if this is indeed the case, we performed a
resampling procedure to try and correct for the differences in stellar properties.
Our source population was the 154 hot Jupiters in the $G < 12.5$ sample described in Section \ref{ssec:selection}.
From this source population, we randomly drew planets to create sets of 41 hot Jupiter hosts (the same number as in the $G < 10.5$ sample) using a rejection sampling procedure, wherein planet hosts were accepted with a probability proportional to the density of \Kepler targets in color-magnitude space.
We generated 1000 resampled catalogs of HJ host stars, each of which had a distribution in color-magnitude space that was statistically indistinguishable from that of the \Kepler hot Jupiter hosts.

Figure \ref{fig:period_rad_comparison} shows that the properties of
the resampled catalogs resemble those of the \Kepler sample more 
closely than the actual magnitude-limited sample.
K-S tests comparing the radius distribution of the
resampled catalog with the \Kepler HJs gave a median $p$-value of 0.13,
indicating that the null hypothesis of identical parent distributions
can no longer be rejected.
This suggests that the apparent difference in the planet radii of the
$G < 10.5$ and \Kepler samples, as discussed above, are at least in
part due to differences in the distribution of host star properties.

Nonetheless, these results must be interpreted with caution, due to the small sizes of the two samples.
Our resampling process showed that due to the small number of planets, features in the period and radius distributions cannot be identified conclusively, especially when considering possible differences in stellar type.
Furthermore, our claim that the $G < 10.5$ magnitude-limited sample is ``possibly complete'' is based only on the total number of detections as a function
of apparent magnitude.
To check if this is really the case, one would need to understand and model the selection function for the surveys for nearby transiting hot Jupiters,
a much larger effort which we did not attempt.
%If the hot Jupiter occurrence rates for the two stellar populations differ (as discussed in the following section), then the magnitude-limited sample may be missing certain hot Jupiters that bias the period and radius distributions.

\subsection{Comparison with Previous Occurrence Rates \label{ssec:prev_occurrence}}
As a check on the matching procedure described in Section \ref{sec:method}, we used
this same procedure to estimate the occurrence rate of hot Jupiters in the \Kepler sample.
We did so by simply matching $\mathcal{S}_K$ with itself.
This yielded many realizations of catalogs of transiting hot Jupiters
that we used to estimate the sampling uncertainty.
To find the total occurrence rate of hot Jupiters, rather than the rate of transiting hot Jupiters, we weighted each transiting planet by the product of $\Rstar/a$ (the inverse transit probability) and the factor $1/0.9$ that accounts for our restriction on the
transit impact parameter.
The result was an estimated occurrence rate of $0.38 \pm 0.06\%$.
This is in line with the estimate of $0.43^{+0.07}_{-0.06}\%$ by \cite{Masuda2017},
who used a similar procedure.
Other studies of the \Kepler hot Jupiter population have also arrived at similar results.
\cite{Santerne2016} found an occurrence rate of $0.47\pm0.08\%$, and
\cite{Petigura2018} found a rate of $0.57^{+0.14}_{-0.12}\%$ based on a more limited subset of the \Kepler stars.

These \Kepler occurrence rates are a factor of $\sim$2 lower than those derived from other studies, albeit with modest statistical significance.
\cite{Cumming2008} found a hot Jupiter occurrence rate of $1.5 \pm 0.6$\% from the Keck radial-velocity surveys, although the parent stellar sample was not constructed blindly with regard to planet host status.
\cite{Wright2012} used a cleaner sample from the Lick and Keck planet searches, finding an occurrence rate of $1.2 \pm 0.4\%$.
This is consistent with the result
of $0.89\pm0.36\%$ reported by \cite{Mayor2011} based on the HARPS and CORALIE radial-velocity searches.
This discrepancy is not limited to comparisons between transit and radial-velocity surveys.
Based on the CoRoT transit survey, \cite{Deleuil2018} reported
a hot Jupiter occurrence rate of $0.98\pm0.26\%$, higher than the \Kepler
results and consistent with the radial-velocity results.

Many authors have investigated possible reasons for the discrepancies in occurrence rates.
\begin{itemize}

    \item \cite{Wang2015} argued that $12.5\pm0.2\%$ of hot Jupiters in the \Kepler sample were likely misidentified as smaller planets due to flux contamination by nearby stars or errors in the estimated stellar radius, but this would be insufficient to account for a factor-of-two difference in measured occurrence rates.
    
    \item \cite{Guo2017} investigated whether the strong association between 
    stellar metallicity and hot Jupiter occurrence could be responsible for the discrepancies. Through spectroscopic observations, they found that \Kepler stars
    are more metal-poor by about $0.04$~dex than the stars in the Lick and Keck radial-velocity surveys, whereas a full resolution of the discrepancy between the point estimates of the hot Jupiter occurrence rates would have required a metallicity difference of 0.2--0.3~dex.
    
    \item \cite{Bouma2018} used simple analytic models to
    argue that the observational biases arising from unrecognized binaries in the \Kepler sample are not large enough to resolve this discrepancy.

    \item \cite{Moe2020} advanced a hypothesis that binarity is responsible for the discrepancies after all, due to an astrophysical effect.
    Hot Jupiters may not be able to form around a star with a close stellar companion ($a < 100$~AU). Because close binaries are systematically excluded from radial-velocity surveys, the inferred occurrence rate of hot Jupiters would be higher than
    in transit surveys that do not exclude close binaries. 
    This would still leave unexplained the relatively high occurrence rate derived from CoRoT survey \citep{Deleuil2018}.

\end{itemize}
If the true occurrence rate of hot Jupiters around FGK stars
is higher than has been inferred from the \Kepler sample, than the matching
procedure we used in this study would underestimate the number of hot Jupiters
that we expect to find around nearby bright stars.
In that case, the ``possibly complete'' sample that we discussed in Section \ref{ssec:complete_samples} would be farther from complete
than it originally appeared.

\section{Summary and Conclusions \label{sec:concusion}}

To better understand the origins of hot Jupiters, it would help to have a better
census of their properties: the distributions and correlations between their periods,
radii, masses, orbital parameters, host star parameters, and occurrence of companion
planets and companion stars.
Currently, the census that is easiest to interpret statistically is the
sample of about 40 hot Jupiters found by the \Kepler mission, which was capable
of detecting transiting hot Jupiters with $\gtrsim\,99\%$ probability
around more than one hundred thousand FGK stars.
The total number of known hot Jupiters is an order of magnitude larger than the
number in the \Kepler sample, but demographic studies cannot yet take full
advantage of this much larger sample size
because the unknown and undoubtedly complex selection functions of the many surveys that have found
hot Jupiters.

We did not attempt to model these selection functions. Instead, based on the simpler
considerations of the distribution of apparent magnitudes and the occurrence of transiting
hot Jupiters in the \Kepler survey, we have shown that the current sample
of transiting hot Jupiters is consistent with being complete down to a limiting
apparent magnitude of 10.5, if we exclude the region of the sky within $10^\circ$ of the galactic
plane.
We examined this subsample of hot Jupiter hosts alongside the \Kepler sample
to compare the distributions of orbital periods and planet radii,
which are indistinguishable, at this stage, after accounting for
differences in the color-magnitude distribution of the host stars.
If there are any differences, or if the longstanding discrepancies between the hot Jupiter occurrence rates measured in different surveys turn out to
have interesting astrophysical origins, then only larger samples of planets will reveal them.

We showed that our current knowledge of hot Jupiter demographics
is far from complete.
We quantified the limiting magnitude and galactic latitudes of stars around which
we need to search for hot Jupiters, finding that many planets remain to be
found even around relatively bright stars, which should be easily detectable
by the TESS, and which are also nearby and bright enough to appear
in many other all-sky surveys, such as the \Gaia astrometric
survey and the APOGEE spectroscopic surveys.
The overlap of these surveys may allow us to discover new connections between hot Jupiters,
the properties of their host stars, and the presence of wide-orbiting companions.

To increase the size of the statistically useful sample of hot Jupiters by an order
of magnitude, we should aim for a complete sample of hot Jupiters down
to a limiting
apparent magnitude of about $G = 12.5$.
Around such stars, we expect $424^{+98}_{-83}$ transiting (and non-grazing)
hot Jupiters, of which
154 are already known, leaving approximately 250 to be discovered and confirmed.
This would represent a significant effort but also a major advance
in our understanding of hot Jupiter origins.
TESS presents an immediate opportunity to perform this task,
thanks to its nearly all-sky coverage, 27-day observing baseline,
and high photometric precision, enabling the construction of
a nearly complete sample of HJs down to 12th magnitude \citep{Sullivan2015,Zhou2019}, while the homogeneous and continuous data will make it possible to understand the selection function much better than those of previous ground-based transit surveys.

\acknowledgements
This research made use of the NASA Exoplanet Archive, which is operated by the California Institute of Technology, under contract with the National Aeronautics and Space Administration under the Exoplanet Exploration Program.
The work also made use of data from the European Space Agency (ESA) mission
{\it Gaia} (\url{https://www.cosmos.esa.int/gaia}), processed by the {\it Gaia}
Data Processing and Analysis Consortium (DPAC,
\url{https://www.cosmos.esa.int/web/gaia/dpac/consortium}). Funding for the DPAC
has been provided by national institutions, in particular the institutions
participating in the {\it Gaia} Multilateral Agreement.
Work by SWY and JNW was funded by the Heising-Simons Foundation and
the TESS project (NASA contract NNG14FC03C).
JH acknowledges support from the TESS GI Program, programs G011103 and G022117, through NASA grants 80NSSC19K0386 and 80NSSC19K1728.

\facility{Exoplanet Archive}
\software{
\texttt{astropy} \citep{Astropy13,Astropy18};
\texttt{numpy} \citep{Numpy};
\texttt{scipy} \citep{Scipy};
\texttt{matplotlib} \citep{Matplotlib}.}

\bibliography{bibliography,software}
\bibliographystyle{aasjournal}

\end{document}